\documentclass{article}

\usepackage{arxiv}

\usepackage[utf8]{inputenc} 
\usepackage[T1]{fontenc}    
\usepackage{hyperref}       
\usepackage{url}            
\usepackage{booktabs}       
\usepackage{amsfonts}       
\usepackage{nicefrac}       
\usepackage{microtype}      
\usepackage{lipsum}		
\usepackage{graphicx}
\usepackage{natbib}
\usepackage{doi}
\usepackage{xcolor}
\usepackage{caption}
\usepackage{subcaption}
\usepackage{amsmath}
\usepackage[ruled,vlined]{algorithm2e}

\title{Tackling Imbalanced Data in Cybersecurity with Transfer Learning: A Case with ROP Payload Detection}

\author{
	Haizhou Wang \\
	College of Information Sciences and Technology\\
	The Pennsylvania State University\\
	University Park, PA, 16802 \\
	\texttt{hjw5074@psu.edu} \\
	\And
	Peng Liu \\
	College of Information Sciences and Technology\\
	The Pennsylvania State University\\
	University Park, PA, 16802 \\
	\texttt{pxl20@psu.edu} \\
}

\date{}

\renewcommand{\headeright}{}
\renewcommand{\undertitle}{}

\hypersetup{
pdftitle={Tackling Imbalanced Data in Cybersecurity with Transfer Learning: A Case with ROP Payload Detection},
pdfsubject={cs},
pdfauthor={Haizhou Wang, Peng Liu},
pdfkeywords={Transfer Learning, Domain Adaptation, Cybersecurity, Return-Oriented Programming, Imbalanced Dataset},
}

\begin{document}
\maketitle

\begin{abstract}
In recent years, deep learning gained proliferating popularity in the cybersecurity application domain, since when being compared to traditional machine learning, it usually involves less human effort, produces better results, and provides better generalizability. However, the imbalanced data issue is very common in cybersecurity, which can substantially deteriorate the performance of the deep learning models. This paper introduces a transfer learning based method to tackle the imbalanced data issue in cybersecurity using Return-Oriented Programming (ROP) payload detection as a case study. We achieved 0.033 average false positive rate, 0.9718 average F1 score and 0.9418 average detection rate on 3 different target domain programs using 2 different source domain programs, with 0 benign training data samples in the target domain. The performance improvement compared to the baseline is a trade-off between false positive rate and detection rate. Using our approach, the number of false positives is reduced by 23.20\%, and as a trade-off, the number of detected malicious samples is reduced by 0.50\%.
\end{abstract}


\section{Introduction} \label{sec:introduction}

Deep learning becomes popular in the fields of cybersecurity in recent years \citep{log1,log2,log3,log4,log5, mem1, mem2,mem3,mem4}, because deep learning based methods perform at least as good as the traditional methods do when enough high-quality data are available, and is more general and cost-effective. However, one of the challenges when applying deep learning to the cybersecurity application domain is the imbalanced data issue, which can deteriorate the performance of the deep learning models. Imbalanced data situations are quite common in cybersecurity. For example, in network intrusion detection, the amount of benign traffic is orders of magnitudes greater than malicious ones. Another example is in tackling insider's threat, where the amount of normal behavior data is orders of magnitudes greater than malicious behavior. In this paper, we present a transfer learning based method to tackle the imbalanced data issue in cybersecurity using Return-Oriented Programming (ROP) payload detection as a case study. ROP is an exploit technique that can be used to perform code reuse attacks (CRA) through the Internet, which is still a prominent exploit technique today used to defeat Data Execution Prevention (DEP), especially on the legacy platforms without Address Space Layout Randomization (ASLR), because ROP payloads contain no code but only addresses. Even if ASLR is deployed, ROP attacks could still be fairly effective. ROP is therefore well-studied and many methods and tools are proposed to detect the ROP attacks.

In recent years, deep learning based ROP detection methods have been proposed, because it could mitigate several limitations of traditional methods. The advantages of using deep learning to detect ROP attacks include: (1) deep learning models can run independently with no overhead on the protected programs; (2) less human heuristics are involved to extract features or patterns; (3) a well trained deep learning model can achieve comparable detection rate (DR) and false positive rate (FPR). To the best of our knowledge, no traditional method has all the advantages mentioned above. For example, defending methods implemented at compiler \citep{onarlioglu2010g} will change the program significantly and may cause runtime overhead; control flow integrity (CFI) based methods \citep{payer2015fine, mashtizadeh2015ccfi} require carefully crafted fine-grained control flow graphs, which could be very challenging; heuristic based methods \citep{chen2009drop, cheng2014ropecker} may suffer from low detection rate.

Deep learning seems promising, but deep learning based ROP detection methods also suffer from the imbalanced data issue just as other cybersecurity subfields. An ROP detection task is essentially a classification task with two classes: benign or malicious. We observe that usually data with one of the two labels could be hard to prepare \citep{zhang2019deepcheck, ropnn}. Therefore, if the time for generating the data is limited, one can choose to either have less total amount of data, or generate less amount of data with hard-to-prepare labels. Since more data is always favored for deep learning, so that the later option is often used, which will lead to imbalanced dataset and neural networks can lose generalizability that can cause the model to be biased to the majority class.

To mitigate the imbalanced data challenge in ROP detection, this paper introduces a transfer learning based solution to mitigate the imbalanced data issue for deep learning based ROP detection methods. The scenario in this paper is based on DeepReturn \cite{ropnn}, which has following assumptions:
\begin{itemize}
    \itemsep-0.1em 
    \item One deep learning model is trained to protect exactly one program.
    \item Enough high-quality data for one program is available.
    \item Extremely imbalanced data for another program is available. There are only very few benign data samples.
\end{itemize}

In our case study, we achieved 0.033 average false positive rate, 0.9718 average F1 score, and 0.9418 average detection rate on 3 different target domain programs using 2 different source domain programs, with 0 benign training sample in the target domain. Compared to the baseline, the number of false positives is reduced by 23.20\%, and as a trade-off, the number of detected malicious samples is reduced by 0.50\%. Our contributions can be summarized as follow:
\begin{itemize}
    \item Propose a new domain adaptation based method to train a cyber-attack detection model using an extremely imbalanced dataset.
    \item Discuss the performance trade-offs of the proposed approach.
    \item Present the insights about how transfer learning helps to achieve the improved results.
\end{itemize}

In the rest of the paper, Section \ref{sec:background} introduces the backgrounds. Section \ref{sec:motivation} elaborates our motivations. Section \ref{sec:method}  describes the methods. Section \ref{sec:evaluation} evaluates the model using experiments, and answers critical questions we have identified.

\section{Background} \label{sec:background}
\subsection{Return-Oriented Programming}
\begin{figure}[ht]
    \centering
    \includegraphics[width=0.95\linewidth]{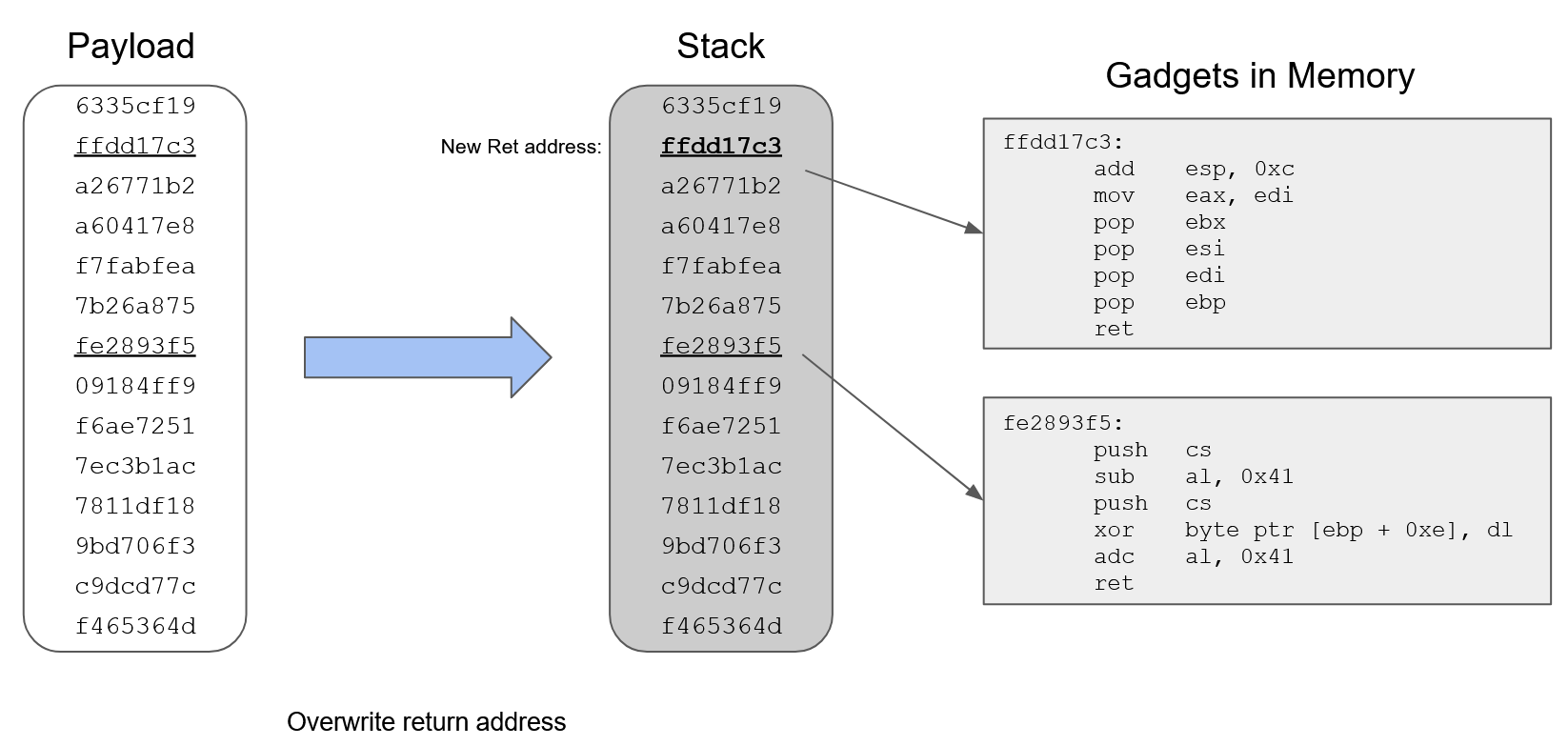}
    \caption{Workflow of ROP Attacks}
    \label{fig:rop_background}
\end{figure}

Return oriented programming (ROP) \citep{rop} and its variants \citep{jitrop, bletsch2011jop, ropwithoutret} are still popular exploit methods today, which provide attackers turing-complete functionalities without inject any code \citep{attack1, attack2, attack3, attack4, ROPonNginx}. The idea of ROP attack is not complicated. The attackers use the instruction sequences that end with an \texttt{ret} instruction to construct the code for their purposes, which are called gadgets. By overwriting the return address of the executing function and loading all the addresses of gadgets needed onto the stack, the attacker will be able to execute a sequence of gadgets, which is called gadget-chain. Figure \ref{fig:rop_background} shows a synthetic example of the ROP exploit process on X86. In this example, the payload arrived at the host and is loaded into a buffer on the stack. This malicious payload segment contains two addresses, \texttt{0xffdd17c3} and \texttt{0xfe2893f5}, which are addresses in the code segment (i.e. \texttt{.text} segment) of the beginning of gadgets. If the address \texttt{0xffdd17c3} overwrites the original return address in the stack frame, it will cause the whole gadget chain to be executed and the program will be exploited by the attacker. Since there are abundant instruction sequences (and thus gadgets) available in the memory when the program is loaded in modern operating systems, virtually ROP is turing-complete programming technique. As a result, detection methods against ROP often do not focus on the payloads; instead they will directly focus on detecting the gadget-chains.

In practice, the most important aspect of an ROP payload is the addresses of the gadgets and their layout. In the simplest scenario, where only \texttt{ret} instruction is used to chain up gadgets, the attacker needs to ensure the value stored in the \texttt{\%esp} register is pointing to the address of the beginning of the next gadget when \texttt{ret} is executed. Usually \texttt{pop} instructions are used to manipulate the \texttt{\%esp} register. If the distance of the addresses of two adjacent gadgets in the memory is 4 bytes, then one \texttt{pop} instruction will be needed. This is also illustrated by the first gadget in Figure \ref{fig:rop_background}. Since it is not common to have many \texttt{pop} instructions in a roll, the addresses of adjacent gadgets are usually not far away in the payload.

\subsection{Traditional ROP Detection Methods}
The majority of the traditional ROP detection methods can be categorized into 2 kinds: heuristic-based, and CFI-based. Heuristic-based methods use heuristics and hard coded rules to find ROP gadgets. DROP \citep{chen2009drop} checks the frequency of executed return or jump instructions. kBouncer \citep{kbounce} and ROPecker \citep{cheng2014ropecker} check indirect branches, and issue an alarm if certain abnormal patterns are found. As mentioned in Section \ref{sec:introduction}, these heuristics could be bypassed if the attackers know them, which result in lower detection rate.

CFI-based methods \citep{CFI00,CFI3, CFI4,CFI6,CFI9} use CFI to assist ROP detection. There are two challenges when using CFI: building accurate fine-grained CFG and causing high overhead on the program.
On one hand, it is shown that building complete and accurate fine-grained CFG is very challenging \citep{burow2017control}, and in fact many works shows that attackers can circumvent CFIs using imperfect CFGs \citep{CFI4, JX2, JX5, ROP11}. On the other hand, CFI may introduce significant overhead to the program\citep{CCFI, payer2015fine}, which is not acceptable for performance critical services.

There are other methods that are neither heuristic-based nor CFI-based. \citet{tanaka2014n} introduced n-ROPDetector, which checks whether a set of function addresses are presented in the payload. Since the method focuses on the payload, attackers can insert obfuscation to avoid being detected. \citet{polychronakis2011rop} proposed an ROP detection method based on speculative code execution, which will issue the alarm if four identified gadgets are executed. However, this could cause a high false positive rate, since normal instruction sequences can contain more than four gadgets, as shown by \citet{stancill2013check}. There are also statistical-learning-based methods \citep{elsabagh2017detecting, pfaff2015learning}, which usually cannot handle large dataset and need handcrafted features.

\subsection{Deep Learning Based ROP Detection Methods}

Deep learning is widely used to solve many security problems, such as log anomaly detection \citep{log1,log2,log3,log4,log5}, memory forensics \citep{mem1, mem2,mem3,mem4}, etc., where data are either widely available or easy to prepare. However, ROP attack detection, or more generally, CRA detection are relatively less popular. It is also observed that preparing the data is the most challenging part for applying deep learning to detect ROP attacks. \citep{ropnn, zhang2019deepcheck,chen2018henet}. For example, \citet{chen2018henet} proposed a unique data representation for traces acquired from Intel PT, which is a 2-dimensional grid data structure that can be used to training neural networks; \citet{zhang2019deepcheck} proposed a specialized fine-grained CFG and a unique way to create malicious data.

Deep learning based ROP detection methods are showing promising results and have two major advantages: 1) usually minimal or no overhead and 2) less human efforts needed to identify heuristics and patterns. Deep learning based ROP detection methods usually have minimal overhead because deep learning models can be deployed separately. Since the deep learning model can capture and extract features automatically, as long as proper representation is provided, no human effort is needed for the pattern finding. However, it seems challenging to prepare the data to train the neural network, and one of the biggest issues is the imbalanced data issue.

\subsection{Transfer Learning in Cybersecurity}

According to the survey by \citet{pan2009survey}, two major categories of transfer learning are inductive transfer learning and transductive transfer learning. Inductive transfer learning focus task knowledge transfer, whereas transductive transfer learning focus data domain (representation) transfer. The most important difference between inductive and transductive transfer learning is whether the label information is available in the target domain. In transfer learning, by convention, the domain where knowledge is transferring from is called source domain, and where knowledge is transferring to is called target domain.

Transfer learning has been widely used in many application domains (i.e. computer vision, natural language processing, etc.), but it is not as popular in cybersecurity. Recently, there are works that use transfer learning to solve the imbalanced data issue in intrusion detection, vulnerability detection, and IoT attack detection. For the intrusion detection, researchers usually use public intrusion detection dataset, which are network packet data. The knowledge transfer usually is cross-exploit, that is transferring the knowledge a model learned to detection one exploit for detecting another exploit. \citet{zhao2017feature} introduced HeTL, which is a non-deep learning transfer learning method that can construct a common representation for source and target domain data using spectral transformation. Then \citet{zhao2019transfer} introduced CeHTL to preprocess the data for HeTL. Based on HeTL, \cite{sameera2020deep} introduced a method that uses manifold alignment to construct common representation instead of spectral transformation.

There are also deep learning based transfer learning for intrusion detection. \citet{gangopadhyay2019domain} introduced a Convolutional Neural Network (CNN) based intrusion detection method, and illustrated that the cross-exploit transfer learning is possible by adding one more layer to the neural network and training on different tasks. \citet{singla2020preparing} proposed an adversarial domain adaptation method, which is inspired by generative adversarial net (GAN) \citep{goodfellow2014generative}.

Transfer learning is also used for vulnerability detection. \citet{nguyen2019deep} proposed a GAN-based domain adaptation method for vulnerability detection at source code level. Another work called CD-VulD \citep{liu2020cd} proposed an easy-to-interpret cross-domain vulnerability detection at source code level. One observation is that most of the vulnerability detection using transfer learning is focusing on source code level, and to some extent, similar to a natural language processing problem. There are also works focusing IoT device attack detection. \citet{vu2020deep} proposed a autoencoder based method to do cross-device knowledge transfer, where the common data representation learning is guided by Mean Maximum Discrepancy (MMD).

There are also other transfer learning applications in cybersecurity. \citet{ampel2020labeling} proposed a transfer learning method to transfer the features learned from publicly available exploit scripts to newly created scripts. \citet{grolman2018transfer} introduced a cross-platform and cross-version encrypted application data analysis method using transfer learning.

\subsection{Domain Adaptation}

Domain adaptation is a subfield of transfer learning, which is used to solve transductive transfer learning problems. One common strategy to do domain adaptation is constructing common representation (i.e. with the same underlying distribution) for source and target domain data. This can be achieved by using a very popular metric for domain adaptation called Mean Maximum Discrepancy (MMD) proposed by \citet{gretton2012kernel}, which can be used to determine if sets of samples are from the same distribution. In other words, a small MMD indicates the samples are from the same distribution. Many researchers found that a neural network can be trained using MMD as a part of the loss function to extract features from data of both domains, so that the features extracted follows the same distribution \citep{long2015learning, tzeng2014deep, rozantsev2018beyond}.

Instead of using MMD based methods, many researchers choose another route: reconstructing the data \citep{ghifary2016deep, zhu2017unpaired}. The essence of this route is to use generative models such as Generative Adversarial Nets (GAN) \citep{goodfellow2014generative} and Variational AutoEncoder (VAE) \citep{kingma2013auto} to reconstruct the data. Using the object detection example in grey and colored scale, one can reconstruct the object in grayscale into colored mode.

\section{Motivation and Problem Statement} \label{sec:motivation}

\subsection{DeepReturn and Data Preparation Process} \label{sec:data_prep_process}

To elaborate our motivation, we first breifly explain the workflow of DeepReturn and its data preparation process. DeepReturn is designed to detect ROP attacks from the network for a single program. To launch an ROP attack, the payloads arrive through the network, and are then sent to the programs that can be exploited. A malicious payload always contains addresses of gadgets in executable segments (i.e \texttt{.text}) of the target program, so by chaining up the gadgets found in the address space of the loaded program, the gadget-chains can be formed and executed. Similarly, regular data arrives through the network too, and it may or may not contain addresses of executable codes. If it does, then one can also chain up a "gadget-like" instruction sequence that may or may not be executable. In DeepReturn, a neural network is used to determine whether an instruction sequence is an actual gadget-chain (malicious) or just a "gadget-like" instruction sequence (benign), and then to determine whether the input data arrived is an ROP payload or just a piece of regular data. Therefore, the training data for the neural network are instruction sequences, where the malicious data are the gadget-chains and the benign data are the "gadget-like" instruction sequences.

\begin{figure}[t]
    \centering
    \includegraphics[width=0.8\linewidth]{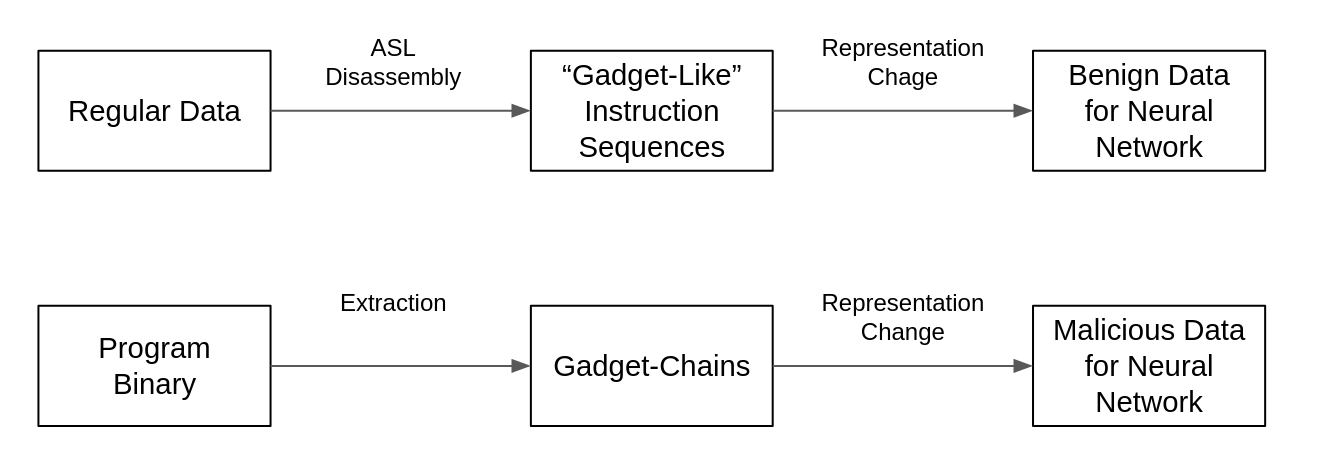}
    \caption{The flow diagram for the data preperation process in DeepReturn}
    \label{fig:data_prep_process}
\end{figure}

The flow diagram of the data preparation process is shown in Figure \ref{fig:data_prep_process}. The authors of DeepReturn, \citet{ropnn} called the process of chaining up the instruction sequences Address Space Layout (ASL) guided disassembly. The detail of the ASL guided disassembly will be explained in Section \ref{sec:data_prep}. As shown in Figure \ref{fig:data_prep_process}, malicious data is prepared by extracting the gadget-chains directly from the binary, and the benign data is generated by chaining up the instruction sequences using ASL guided disassembly. During the benign data generation, a piece of regular input data may or may not contain addresses of executable codes, and therefore, not all input regular data can be used to form "gadget-like" instruction sequences. In DeepReturn, authors found that the input data which contains addresses of executable code so that it can form "gadget-like" instruction sequences is very rare, causing the cost of generating benign data samples extremely expensive. In contrast, malicious data does not have to be generated through an actual payload. Instead, malicious samples can be easily generated by using gadget-chain generating tools, such as ROPGadget \citep{ROPgadget2015}.

\subsection{Issue of Imbalanced Data}
As shown in Section \ref{sec:data_prep_process}, it is quick and cheap to generate malicious data samples; however, it is very expensive to generate benign data samples. For example, in DeepReturn, it takes 7 hours to generate benign data on a cluster node with 96 CPUs for web server programs and FTP server programs. In other words, whenever the model needs to be trained or retrained, a cluster node is needed and kept running for 7 hours before the training phase. In large-scale scenarios, the deep learning based method becomes less practical, because there are many programs that can suffer from ROP attacks so that many models need to be trained. Besides, it is widely agreed that programs should be kept updated for security patches, so the number of training sessions will further increase.

We believe the imbalanced data is a real-world issue that may cause many deep learning based solutions impractical. The essence of the imbalanced data issue is the trade-off between cost and security. In the simplest scenario, the model maintainer can choose to train the model with the imbalanced dataset, which can cause the model to be biased. Depending on the level of imbalance, the model performance can vary. This is the trade-off between the level of imbalance and the model performance, which essentially is the trade-off between cost and security: choosing imbalanced data with higher level will cause the system to be inadequately protected, whereas choosing to produce more balanced data can increase the cost significantly.

In essence, solving the imbalanced data issue can avoid such difficult cost vs. security trade-off. In case of the DeepReturn, if the time to generate benign data is not 7 hours on a cluster node but 1 hour on a personal computer, the approach will become much more practical and scalable. This can be achieved either by reducing the data needed or generating the data quicker, because what matters here is not the number of data, but the time needed to generate the data. Usually, reducing the data generation time is difficult, so that the only practical way to reduce the time is to generate less samples. Therefore, whether the approach is practical largely depends on whether the performance can be maintained with the significantly reduced amount of data.

\subsection{Problem Statement}
In order to address the scalability issue of DeepReturn, and to make the deep learning based approaches more practical in real world, we aim to solve the following problem:

Many programs may suffer from ROP attacks if there exists a vulnerability that can be used to overwrite the return address of a function to arbitrary value. It is shown that deep learning can be leveraged to detect ROP attacks effectively, providing high detection rate and very low false positive rate. However, deep learning based methods often suffer from the imbalanced data issue when used to detect ROP attacks. To mitigate the effect of the imbalanced data, transfer learning may be leveraged to improve the performance of a deep learning model. The problem is whether transfer learning could be used to make the deep learning based approaches effective in the presence of imbalanced data, scalable, and significantly more practical.

\section{Method} \label{sec:method}

\subsection{ASL-Guided Disassembly} \label{sec:data_prep}

\begin{figure}[t]
    \centering
    \includegraphics[width=\linewidth]{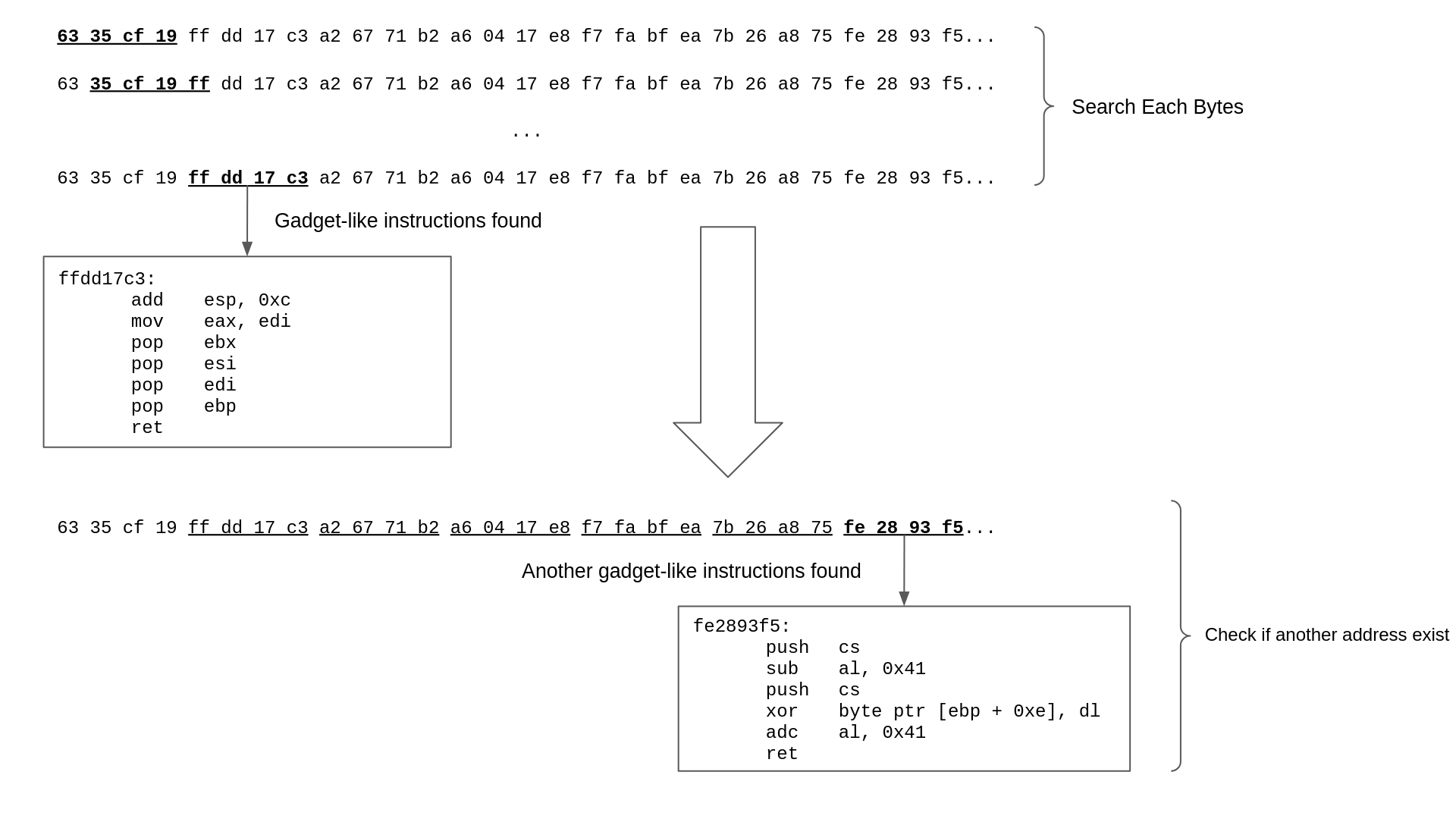}
    \caption{An illustration for ASL-guided disassembly}
    \label{fig:ASL_disassembly}
\end{figure}

To generate benign data for the training phase and extract instruction sequences from the incoming network data during the production phase, DeepReturn uses an approach called ASL-guided disassembly. This section summarizes the process of the ASL-guided disassembly proposed by \citet{ropnn}.

First, the reassembled network data is scanned so that the starting address of potential gadget-chains can be identified. Each byte could be the beginning of an address, and 4 consecutive bytes will be considered as an address (for x86). If an address at $n$ is the start of an executable instruction sequences that end with an indirect branch, then the next 5 to 10 4-byte-long data (i.e. $n+4$, $n+8$...) will be evaluated to see if they are also addresses for such instruction sequences. If yes, then these instruction sequences will be chained up and let the deep learning model decide whether or not it is a gadget chain. Figure \ref{fig:ASL_disassembly} illustrates an example of ASL-guided disassembly. The first step is to find a valid address that points to a potential gadget by searching through the data byte-by-byte, starting at \texttt{0x6335cf19}. The first valid address is found at the byte 4, which is \texttt{0xffdd17c3}. After this address is confirmed to be a gadget address, then we check if another gadget address can be found. In x86, the address takes 4 bytes, so we check the next 5 to 10 4-byte segments. Here another gadget address is found, which is \texttt{0xfe2893f5}, so a data sample is identified.

To reduce the cost in this paper, for programs that need to serve as source domain programs, both benign and malicious data are prepared; for programs only used as target domain programs, only malicious data and few number of benign data samples for validation and testing are prepared.

\subsection{Basic Neural Network Architecture} \label{sec:nn_arch}
In DeepReturn, despite of the sequence data (i.e instruction sequences), \citet{ropnn} shows the Convolutional Neural Network (CNN) performs at least as well as the Recurrent Neural Network (RNN) does, but CNN is much easier to train. In this paper, we have no motivation to change the architecture to a different one, so the classifier used is CNN. For sequence data, 1-dimentional CNN is appropriate. To perform domain adaptation, there are modifications in the fully connected layers based on the original architecture in DeepReturn. The details of the modification are explained in Section \ref{sec:dda}.

The input data are instruction sequences in binary form. After the gadgets and the gadget-like instruction sequences are identified, they will be assembled back to binary. Therefore, for the neural network, the inputs are essentially byte sequences. The atomic unit is the byte, which is represented as an integer number between 0 to 255.

\begin{figure}[ht]
    \centering
    \includegraphics[width=0.95\linewidth]{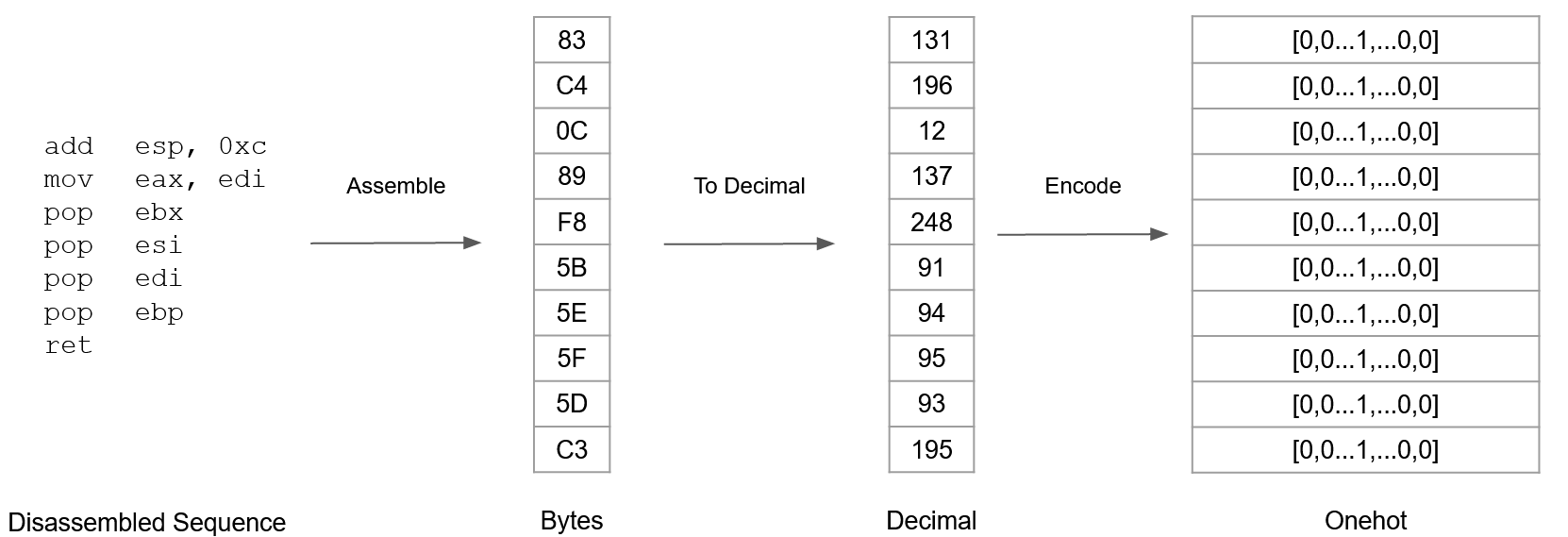}
    \caption{Input Data Representation}
    \label{fig:byte_to_onehot}
\end{figure}

To eliminate the effect of the numerical relationships between each byte (i.e. 255 is larger than 0), one-hot encoding will be used to vectorize each byte. Therefore, the final input data that is ready to feed into the CNN will be a sequence of one-hot vectors. Figure \ref{fig:byte_to_onehot} illustrates how binary instruction sequences are processed after being identified. For example, \texttt{add esp, 0xc} will be assembled to \texttt{0x83 0xc4 0x0c}. Then, it will be transformed to decimal numbers, which is $ [131, 196, 12]$. Finally, the decimal numbers will be encoded to onehot.

The hidden layers follow regular CNN classifier design. Formally, let $F(\mathbf{X})$ be the CNN classifier, and $\hat{y}$, where $0 < \hat{y} < 1 $, be the output of the CNN. Input $\mathbf{X}$ has a shape of $(N, t, s)$, where N is the batch size for mini-batch training, or the dataset size for full batch training; $t$ is is number of elements in the sequence, and $s$ is the feature space size. Here, since onehot vectors are used, and there are 256 unique bytes, the feature space size $s = 256$. We observe that most of the gadget chains and instruction sequences are not longer than 128 bytes, so here $t=128$. The output $\hat{y}$ is the predicted probability of label $y=1$, where $y=1$ means the input is a malicious gadget chain, and $y=0$ means the input is benign gadget-like instruction sequences.

$F(\mathbf{X})$ can be trained by applying gradient descent on cross entropy loss. For binary classification, cross entropy loss is shown in Equation  \ref{eq:cross_entropy}.
\begin{equation} \label{eq:cross_entropy}
    J(\theta) = - y \log (\hat{y}) - (1 - y) \log (1 - \hat{y})
\end{equation}

Overfitting issues are addressed by using dropout and early stopping. The dropout rate is 0.5, and validation data are used to stop the training early. Batch normalization is also used to stabilize the training.

\begin{table}[ht]
    \captionsetup{width=10cm}
    \captionsetup{justification=centering}
	\caption{Parameters of Layers in Basic Neural Network Architecture}
	\centering
	\begin{tabular}{lll}
	\toprule
    Layer Name       & Parameters                                             & Activation \\
    \midrule
    1D Convolutional 1 & 7 x 1 Kernel, 64 Channels, BatchNorm, 0.5 Dropout Rate & ReLU       \\
    1D Convolutional 2 & 5 x 1 Kernel, 64 Channels, BatchNorm, 0.5 Dropout Rate & ReLU       \\
    1D Convolutional 3 & 3 x 1 Kernel, 64 Channels, BatchNorm, 0.5 Dropout Rate & ReLU       \\
    Fully Connected 1  & 256 Neurons, BatchNorm                                 & ReLU       \\
    Fully Connected 2  & 1 Neuron, Output Layer                                 & None    \\
    \bottomrule
    \end{tabular}
	\label{tab:model_arch}
\end{table}

The details of the neural network architecture is shown in Table \ref{tab:model_arch}. Notice that we do not use the exact same architecture as DeepReturn does, because we need extra layers to conduct Domain Adaptation, which will be explained in detail in Section \ref{sec:dda}.

\subsection{Deep Domain Adaptation Using Mean Maximum Discrepancy} \label{sec:dda}

Existing methods to solve the class imbalance issue have two major categories: data-based \citep{hensman2015impact,lee2016plankton,pouyanfar2018dynamic} and model-based \citep{wang2016training}. The effect of the imbalanced data is minimized by sampling the data in dedicated ways in data-based methods, and by changing model architectures and training processes in model-based methods. In our scenario, data-based methods cannot utilize one important advantage we have: some high quality datasets for different programs are available. Therefore, we choose a model-based method: transfer learning, to tackle the imbalanced data issue.

The major difference between detecting ROP attacks for one program from detecting ROP attacks for another program is that the instructions available are different. Therefore, the resulting data representation will be different. Based on this observation, a subfield of transfer learning, domain adaptation fits our task very well, because domain adaptation solves transductive transfer learning problems where the data domains are different, but the tasks are the same.

Many popular domain adaptation methods are based on Mean Maximum Discrepancy (MMD), which is introduced by \citet{gretton2012kernel}. MMD can be used as a distance between two distributions, given samples retrieved from each distribution. Formally, MMD is defined in reproducing kernel Hilbert space (RKHS), denoted as $H$. Let feature extraction layers of our neural network be $f_{\theta}(x)$, where $\theta$ are model parameters. Then, given two random variables $X$ and $Y$ drawn from distribution $p$ and $q$, respectively, the MMD is defined as:
\begin{equation} \label{eq:mmd}
    \text{MMD} (f_{\theta}, p, q) = \|\mathop{\mathbb{E}}_x [f_{\theta}(X)] - \mathop{\mathbb{E}}_y [f_{\theta}(Y)] \|_H
\end{equation}
Here for $\mathop{\mathbb{E}}_x [f_{\theta}(X)]$ and $\mathop{\mathbb{E}}_y [f_{\theta}(Y)]$, we use Monte Carlo estimation, so that $\mathop{\mathbb{E}}_x [f_{\theta}(X)] = \frac{1}{m} \sum_{i=0}^{m} k(\cdot ,f_{\theta}(x_i))$. The kernel $k$ used is the Gaussian kernel, which is defined as:
\begin{equation}
    k(\boldsymbol{x}, \boldsymbol{y}) = \exp \left(-\frac{\|\boldsymbol{x}-\boldsymbol{y}\|^{2}}{2 \sigma^{2}}\right)
\end{equation}

\begin{figure}[t]
    \centering
    \includegraphics[width=0.65\linewidth]{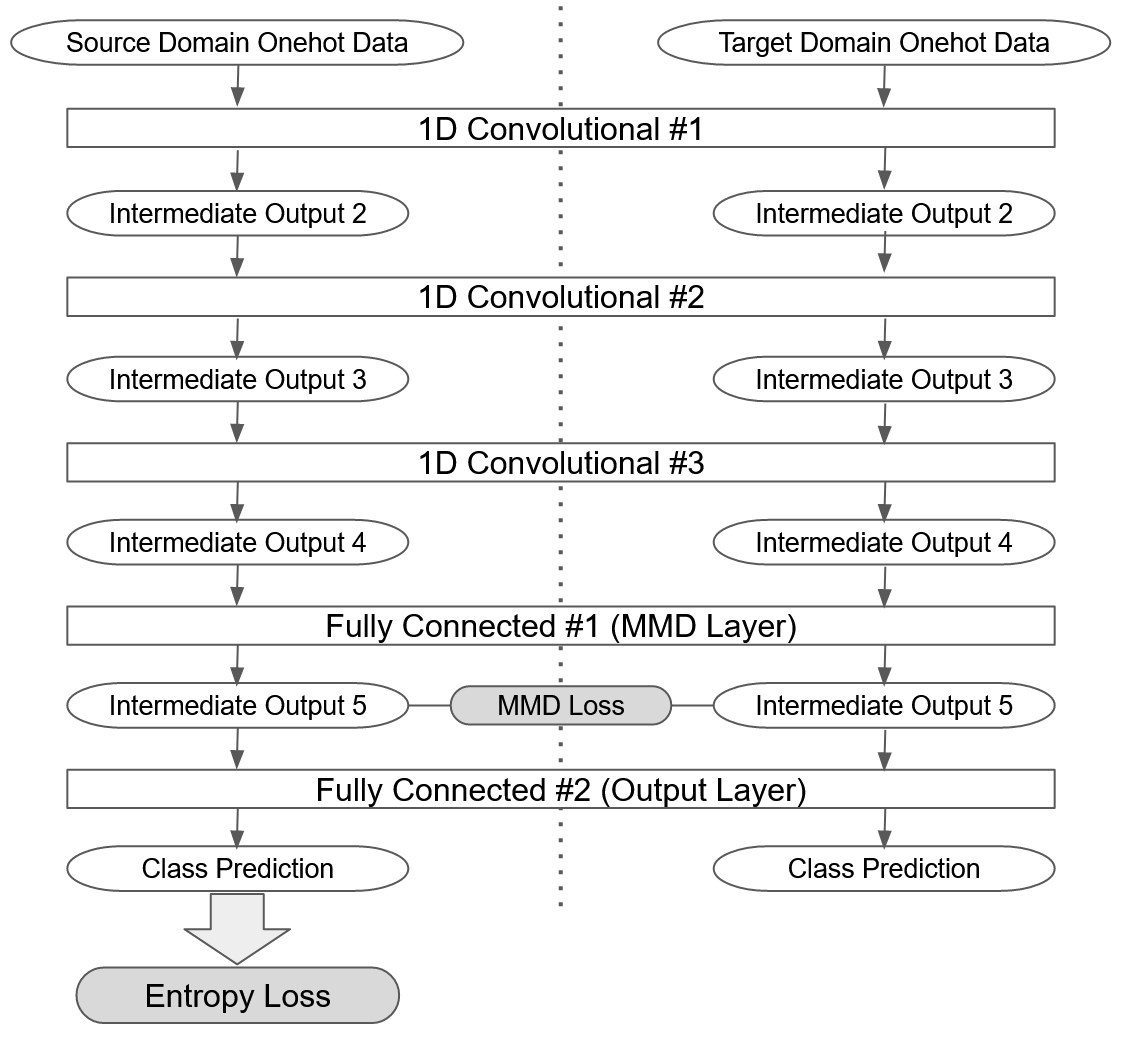}
    \caption{Architecture of the Deep Domain Adaptation Model}
    \label{fig:ddc_arch}
\end{figure}

We first formally define our deep domain adaptation model, and then illustrate the architecture in Figure \ref{fig:ddc_arch}. Let the source domain data to be $X$, and target domain data to be $Y$, MMD then can be found using Equation \ref{eq:mmd}, which will be one part of the loss function. The minimization of MMD ensures that $f_{\theta}(X)$ and $f_{\theta}(Y)$ have the same underlying distribution. The other part of the loss function will be the regular entropy loss. To calculate the entropy loss, extra layers after the $f_{\theta}(X)$ and $f_{\theta}(Y)$ are added. Let the extra layers to be $g_{\theta'}$, then using all data samples $Z$ in both domain, where $Z = {X} \cup {Y}$, the cross entropy loss can be constructed as described in Section \ref{sec:nn_arch}.

As shown in Figure \ref{fig:ddc_arch}, both source and target domain data will be fed into the model. The class prediction for the source domain data will be used to construct entropy loss. During the training phase, the output of the MMD layer, intermediate output 5, will be used to calculate the MMD loss. To get the MMD loss, both source domain and target domain data will need to be fed into the network in one step. The MMD loss will be the calculated MMD between intermediate output of source domain data and the one of target domain data.

\subsection{Training Using No Benign Data in Target Domain}
An important situation is that there are very few benign data in the target domain. In other words, the target domain is extremely imbalanced, and in fact, there is no benign target domain training data. Directly using MMD and entropy loss together is not appropriate, because the classification will be biased to the majority class and the model will lose generalizability. Also, since the label information in the target domain is known, such information should be utilized. The core idea is that, for each epoch,the entropy loss will be first be calculated and minimized using balanced data, and then the MMD will be calculated and minimized using only malicious data in both domains. The benefits for doing so include that: 1) the model will not be biased to any class for the classification task, and 2) it is more appropriate that the MMD is only calculated between malicious data samples in two domains, since there are no benign samples in the target domain.

To prevent overfitting and achieve the best test accuracy, we use early stopping, which requires validation data. The validation dataset contains benign target domain data and this is the only place that requires benign target domain data.  We emphasize the importance of validation data here for early stopping to prevent overfitting, and discuss the number of validation data samples needed in Section \ref{sec:evaluation}. Algorithm \ref{algo:train_loop} summarize the training loop:

\begin{algorithm}[H] \label{algo:train_loop}
\SetAlgoLined
\KwResult{Trained model that can perform ROP detection in target domain}
 Initialize bottom feature extraction layers $f_{\theta}$, top task layers $g_{\theta'}$\;
 Initialize maximum epoch $E$, current epoch $e = 0$\;
 Initialize best models $f_{\theta}^{best}$, $g_{\theta'}^{best}$ and best accuracy $acc^{best}=0$\;
 \While{$e < E$}{
  Update $f_{\theta}$ and $g_{\theta'}$ using Eq. \ref{eq:cross_entropy} and balanced data from source domain\;
  Update $f_{\theta}$ using Eq. \ref{eq:mmd} and malicious data samples from both domains\;
  Validate the model and get the validation accuracy $acc$\;
  \If{$acc > acc^{best}$}{
   $f_{\theta}^{best}=f_{\theta}$\;
   $g_{\theta'}^{best}=g_{\theta'}$\;
   $acc^{best}=acc$\;
   }
   $e = e + 1$;
 }
 \caption{Customized Training Loop for Imbalanced Data in Target Domain}
\end{algorithm}

In our experiments, we use Adam optimizer \citep{kingma2014adam} with a learning rate of 0.001; the maximum epoch $E$ is 25, and the batch size is 32.

\section{Evaluation} \label{sec:evaluation}

To evaluate our method, the baseline is defined as the performance of a DeepReturn model trained using one program performing ROP detection tasks on a different program. To make the comparison fair, the DeepReturn model architecture and the training hyper-parameters are modified to be consistent with our methods.

We first introduce the dataset. Table \ref{tab:data_num} summarizes the number of data samples used during training, validation and testing.

\begin{table}[ht]
    \captionsetup{width=10cm}
    \captionsetup{justification=centering}
	\caption{Number of Data Samples Used During Training, Validation and Testing.}
	\centering
	\begin{tabular}{cccc}
	\toprule
                                         & Train & Validation & Test \\
    \midrule
    Number of Malicious Samples (Source Domain) & 20000 & -          & -    \\
    Number of Benign Samples (Source Domain)    & 20000 & -          & -    \\
    Number of Malicious Samples (Target Domain) & 20000 & 1750       & 7500 \\
    Number of Benign Samples (Target Domain)    & -     & 1750       & 1200 \\
    \bottomrule
    \end{tabular}
	\label{tab:data_num}
\end{table}

To generate benign data, 2 TB PDF documents image data are used as inputs for source domain programs. In the experiment in this paper, there are 20000 benign data samples and 20000 malicious data samples for source domain programs; there are 20000 malicious data samples for target domain. For validation, there are 1750 benign and 1750 malicious target domain program data samples available. Then for the test, there are 1200 benign and 7500 malicious target domain program data samples. Note that for programs only used as a target domain program, only a very few number of benign data samples for validation and testing are prepared.

Accuracy is not selected as one of the performance metrics, because the test data in the target domain is extremely imbalanced. The major reason why the test data is imbalanced is because generating benign data is too expensive, but we want our test dataset to be large. Instead, we use F1 score, false positive rate (FPR), and detection rate (DR). FPR is important because in production, the neural network will likely see more malicious data samples so that the effect of FPR will be amplified. Also, false positives are one of the most important concerns in the industry for cyber-attack detection systems.

We use 4 Internet service programs to evaluate our method, with the following metrics: false positive rate (FPR), f1-score (F1), and detection rate (DR). The 4 programs are proftpd 1.3.0a, vsftpd 3.03, nginx 1.4.0 and Apache httpd 2.2.18. Only proftpd 1.3.0a and vsftpd 3.03 are used as source domain programs.

In the following subsections, we will answer following research questions:
\begin{enumerate}
    \item Can our model provide performance improvement, comparing to source domain model?
    \item What is the minimum amount of validation data needed?
    \item How is the knowledge being transferred?
\end{enumerate}

\subsection{Can our model provide performance improvement, comparing to source domain model?} \label{sec:improvement}

To answer this question, it is important to understand why it is not appropriate to use a model trained using one program to detect ROP attacks for another program. Let program A to be the source domain program, and program B to be the target domain program. Even if a deep learning model is very well-trained to detect ROP attacks for program A using the training data generated from program A, we cannot conclude that this deep learning model will perform as well on program B, because program B may contain instruction sequences that program A may not contain. Besides, as we have no control what features a deep learning model will learn, features that extracted by the deep learning model from program A data may not be appropriate in the context of program B. In other words, deep learning model may conclude that some instruction sequence snippets could be an indicator of an ROP gadget chain, but in fact that is a very specific case for Program A, which could lead to false positive. The results are shown in Table \ref{tab:raw_data}, which has following columns:
\begin{itemize}
    \item Source: Source domain programs. Data for these programs are balanced and sufficient, but benign data are very expensive to generate.
    \item Target: Target domain programs. No benign data are available for training. Only limited amount of benign data for validation.
    \item FPR(A): Baseline false positive rate. Model are trained using data of source domain programs to detect ROP attacks for target domain programs.
    \item FPR(B): False positive rate of our model. Model are trained with domain adaptation.
    \item F1(A): Baseline F1 score.
    \item F1(B): F1 score of our model. 
    \item DR(A): Baseline detection rate
    \item DR(B): Detection rate of our model.
\end{itemize}

As shown in Table \ref{tab:raw_data}, when transfer learning is not used, the false positive rate is high on average. The performance metrics that have improvement with respect to the baseline are in bold. Among the six scenarios, there are 3 cases where the F1 score improved, 4 cases where the FPR improved, and 3 cases where the detection rate improved.

\begin{table}[ht]
    \captionsetup{width=10cm}
    \captionsetup{justification=centering}
	\caption{Performance of the ROP Detection on Target Domain Programs using Source Domain Model and Domain Adaptation Model}
	\centering
	\begin{tabular}{cccccccc}
	\toprule
    Source  & Target  & FPR(A) & FPR(B) & F1(A)  & F1(B)  & DR (A) & DR (B) \\
    \midrule
    vsftpd  & nginx   & 0.0442 & \textbf{0.0375} & 0.9601 & \textbf{0.9657} & 0.9374 & \textbf{0.9458} \\
    vsftpd  & httpd   & 0.0283 & 0.0333 & 0.9508 & \textbf{0.9601} & 0.9151  & \textbf{0.9339} \\
    vsftpd  & proftpd & 0.0708 & \textbf{0.0433} & 0.9733 & 0.9452 & 0.9713  & 0.9096 \\
    proftpd & nginx   & 0.0217 & \textbf{0.0192} & 0.9941 & 0.9881 & 0.9957  & 0.9829 \\
    proftpd & httpd   & 0.0167 & 0.0183 & 0.9923 & 0.9843 & 0.9904  & 0.9754 \\
    proftpd & vsftpd  & 0.0392 & \textbf{0.0125} & 0.9484 & \textbf{0.9709} & 0.9142  & \textbf{0.9475}  \\
    \bottomrule
    \end{tabular}
	\label{tab:raw_data}
\end{table}

The best result achieved is when using proftpd as the source domain program and vsftpd as the target domain program. The improvement of the FPR is from 0.0392 to 0.0125 and the detection rate is from 0.9142 to 0.9475. Meanwhile, we also observe cases where the performance is not improved, such as when vsftpd is the target domain program and proftpd is the source domain program, where the detection rate dropped from 0.9713 to 0.9096. Also on average, it is shown that the number of the false positive results is 23.20\% lower, which is an improvement; and the number of detected positive results (ROP attacks) is 0.50\% lower, which is a deterioration. One observation is that whenever proftpd is used as a source domain program, the performance is already very good without using transfer learning. Intuitively, it may be because proftpd data includes the most useful features that can be captured by a deep learning model for ROP detection for other programs. In contrast, when vsftpd is used as a source domain program, the domain adaptation seems effective and improves the performance.

From the result, we argue that our method can significantly improve the false positive rate with a small trade-off on the detection rate. Regarding the research problem we are trying to answer, the performance of the model is largely depending on the programs in both domains. In cases where the source domain program can provide effective features for ROP detection for target domain program, our model may be less effective; however, whenever the model trained using source domain program performs poorly on target domain program, our model performs well. In most cases, the false positive rates are significantly lower, because more malicious data in the target domain are exposed to the model.

\subsection{What is the minimum amount of validation data needed?}

\begin{figure}
     \centering
     
     \begin{subfigure}[b]{0.45\textwidth}
         \centering
         \includegraphics[width=\textwidth]{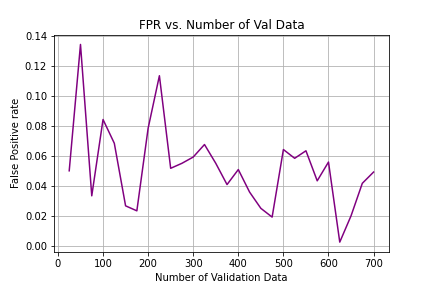}
         \caption{False Positive Rate VS. Number of Validation Data}
         \label{fig:fpr_val}
     \end{subfigure}
     \hfill
     \begin{subfigure}[b]{0.45\textwidth}
         \centering
         \includegraphics[width=\textwidth]{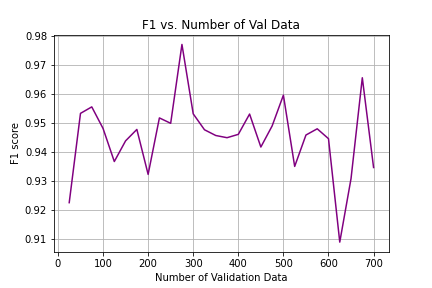}
         \caption{F1 VS. Number of Validation Data}
         \label{fig:f1_val}
     \end{subfigure}

    \caption{Performance VS. Number of Validation Data}
    \label{fig:val_num}
\end{figure}

Generating benign data is expensive, and should be avoided as much as possible, as explained in Section \ref{sec:data_prep_process}. Therefore, one important consideration is the number of validation data needed during the training phase. Similar to training data, validation data is also preferably balanced so that the best model could be selected. Therefore, small amounts of benign target domain data samples are needed for validation.

Extra experiments are conducted to find an appropriate amount of validation data needed. The source domain program is proftpd and target domain program is vsfptd. The result is shown in Figure \ref{fig:val_num}. Figure \ref{fig:fpr_val} shows that the test FPR has a decreasing trend with an increased number of validation data. However, Figure \ref{fig:f1_val} does not show a clear trend for F1. Within 100 validation data, FPR could be as high as 0.13; however, after increasing the number of validation data to over 600, the highest FPR is only about 0.03. In contrast, the F1 score does not change much as the validation data increases, which is about 0.97 all time.

It is expected that FPR will show a clear trend, because as discussed in Section \ref{sec:improvement}, our model improves FPR more than other metrics. However, the trend is far from significant and it is shown that although the more of the validation data, the better the model performance is, the improvement is very limited so that our model can still perform reasonably well with limited amount of validation data.

It is important to point out that needing few validation data does not mean no validation data needed at all. In fact, from our experiment, it is extremely important to have validation data and early stopping during the training phase. The MMD loss is very vulnerable to overfitting, and can result in very bad test performance.

\subsection{How is the knowledge being transferred?}

An interesting question is whether the knowledge is actually transferred, and how the knowledge is transferred. Extra experiments are conducted with proftpd as the source domain program and vsfptd as the target domain program.

\begin{figure}[h]
     \centering
     
     \includegraphics[width=0.55\textwidth]{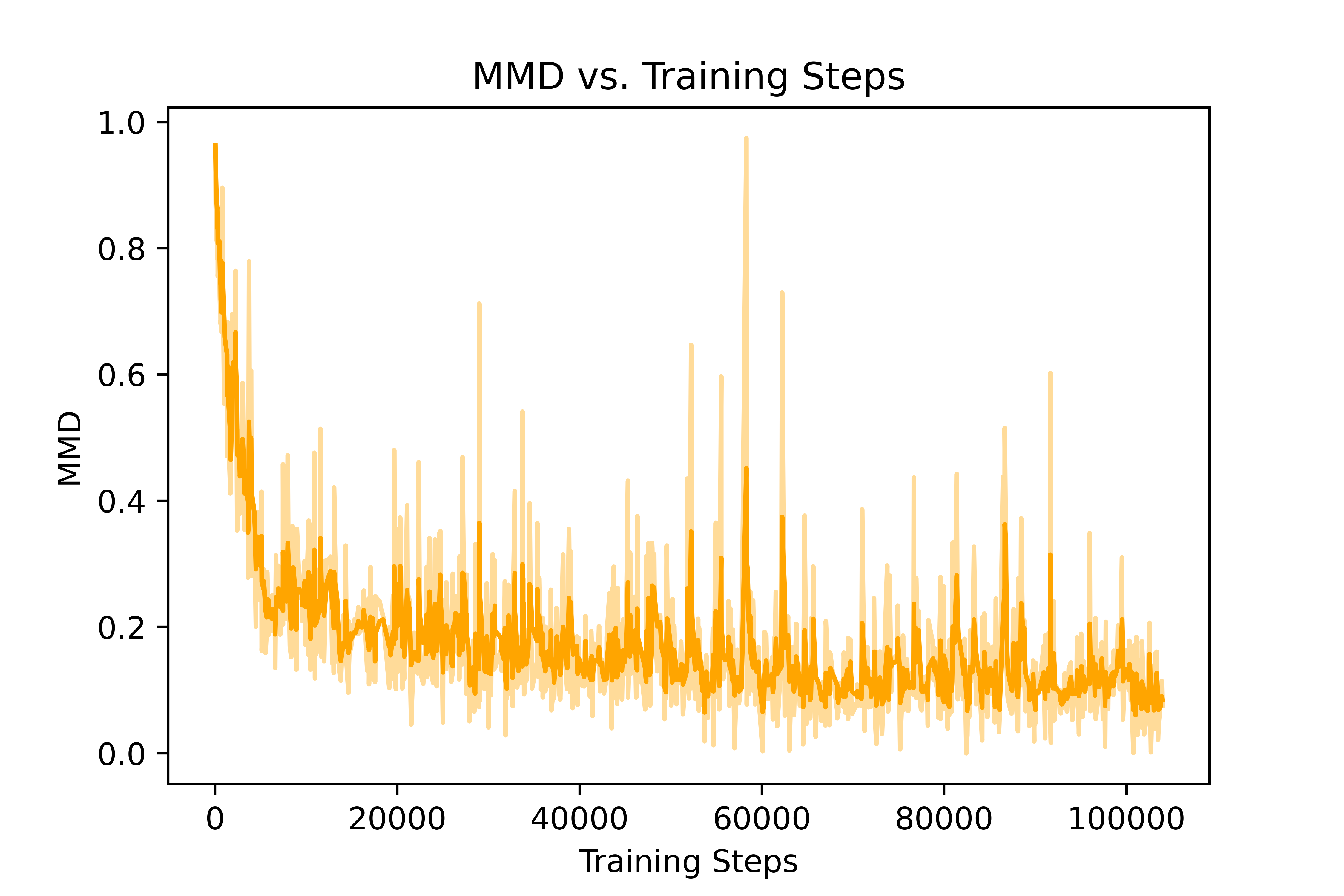}

    \caption{Performance VS. Number of Validation Data}
    \label{fig:mmd_plot}
\end{figure}

Starting with a machine learning perspective, one important factor to consider is the MMD value. Remember MMD can be used as a distance metric for distributions, so whether the average MMD is changed is an indicator whether the source domain data and target domain data are encoded into a similar representation. Figure \ref{fig:mmd_plot} shows the MMD values at different training stages. The plot is smoothed using exponential moving average with $\alpha=0.6$, and the faded lines in Figure \ref{fig:mmd_plot} is the MMD values before smoothing. It is shown that the MMD stops decreasing significantly after 40000 steps. In the beginning the MMD is about 0.9, and after training the MMD is about 0.008. It is clear that the encoded representation of data (i.e. the intermediate output of the layer where MMD loss is formed) from two different domains becomes similar as the training progress.

Next let us dig deeper into this question. We first propose two hypotheses:
\begin{enumerate}
    \item[\textbf{H1:}] Transfer learning helps the model to capture knowledge in target domain and discard features that are not shared by two domains.
    \item[\textbf{H2:}] Transfer learning will not make the model discard source domain knowledge that is useful.
\end{enumerate}

\begin{figure}[t]
    \captionsetup{width=10cm}
    \captionsetup{justification=centering}
    \centering
    \includegraphics[width=0.3\textwidth]{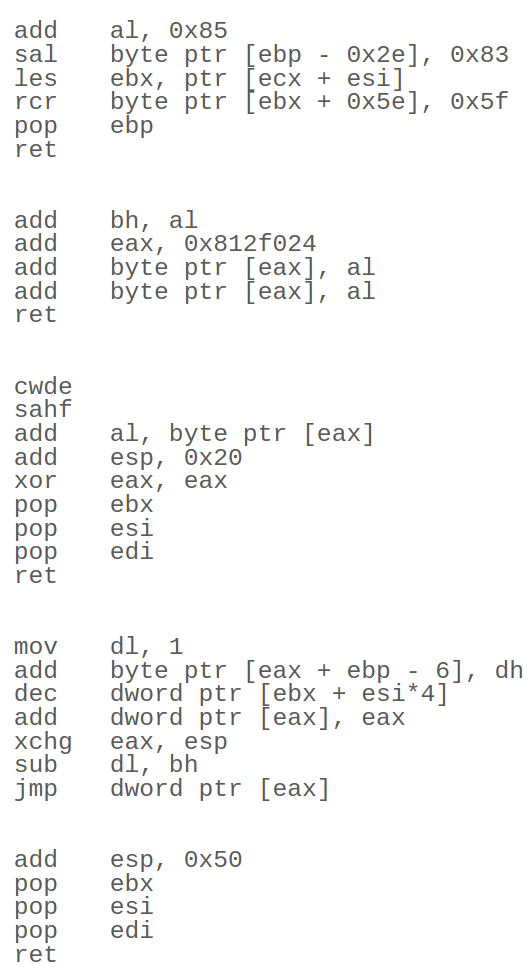}
    \caption{Selected Sample for H1. A Target Domain Malicious Sample that is Correctly Classified by Transfer Learning Model.}
    \label{fig:h1_showcase}
\end{figure}

To verify \textbf{H1}, a sample from the target domain that is correctly classified by model trained using our method, and incorrectly classified by the original model without transfer learning is found. Figure \ref{fig:h1_showcase} is the disassembly of the selected sample. By evaluating the semantic meaning of this gadget-chain snippet, it contains many gadgets for manipulating the stack for jumping to other gadgets (Write-What-Where gadgets), which could be very program-specific because of the different address space layout for different programs. Since this gadget-chain is target-domain-specific, it is not very surprising that the original model, which is completely trained inside the source domain, incorrectly classified it.

We also evaluate the uniqueness of the gadget chain quantitatively by calculating the distance between the instruction sequences using Longest Common Subsequence (LCS) of opcodes. We first find the baseline by calculating the combination pairwise average LCS between source domain and target domain instruction sequences, which is 18.35; then we find the average LCS of the sample in Figure 8 and all other data in source domain, which is 19.42. From the result, we conclude that the selected sample shown in Figure \ref{fig:h1_showcase} is target domain specific, and we expect the extracted feature from this example using our transfer learning model and baseline model should be more different than average. The intuition behind is that target domain special cases should be treated specially, and our model with transfer learning will capture different features to make the classification correct.

The similarity between the extracted features are measured by calculating the distance between the intermediate outputs from two models. Since the baseline model and our model is trained differently, it is not appropriate to make direct comparison between the intermediate outputs from two models, since they have different underlying distributions. To circumvent this issue, we first estimate the distance between two intermediate output spaces as baseline by averaging the combination pairwise distances of all intermediate outputs from both domains, which turns out to be 1.26 using euclidean distance. Then the average distance between the intermediate output of the selected sample and all source domain samples is calculated, which is 1.38. It is shown that compared to the most of the other samples in target domain, the intermediate output of the selected target domain sample is more different (or unique) from the intermediate outputs of source domain.

\begin{figure}[ht]
     \captionsetup{width=10cm}
    \captionsetup{justification=centering}
     \centering
     
     \begin{subfigure}[b]{0.35\textwidth}
         \captionsetup{width=4cm}
        \captionsetup{justification=centering}
         \centering
         \includegraphics[width=\textwidth]{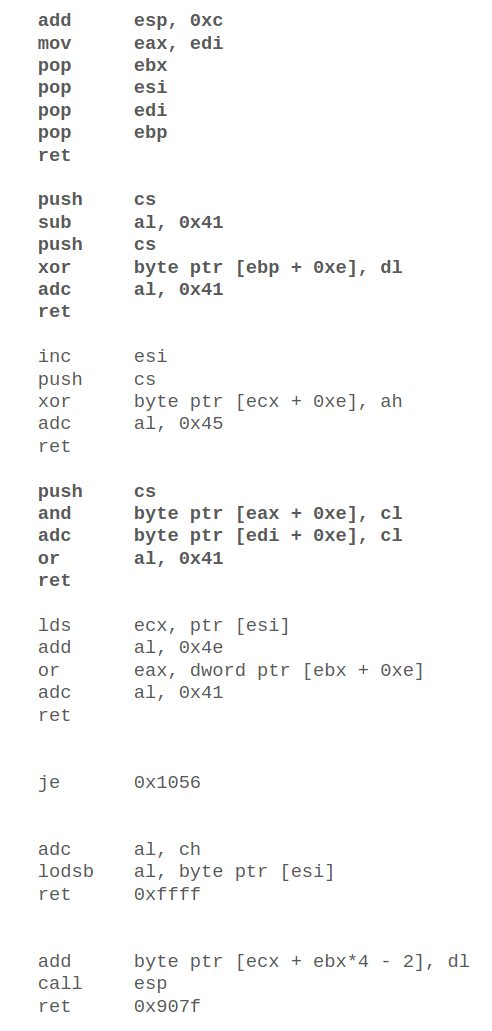}
         \caption{Selected Sample From Source Domain (proftpd)}
         \label{fig:h2_showcase_1}
     \end{subfigure}
     \begin{subfigure}[b]{0.35\textwidth}
         \captionsetup{width=4cm}
        \captionsetup{justification=centering}
         \centering
         \includegraphics[width=\textwidth]{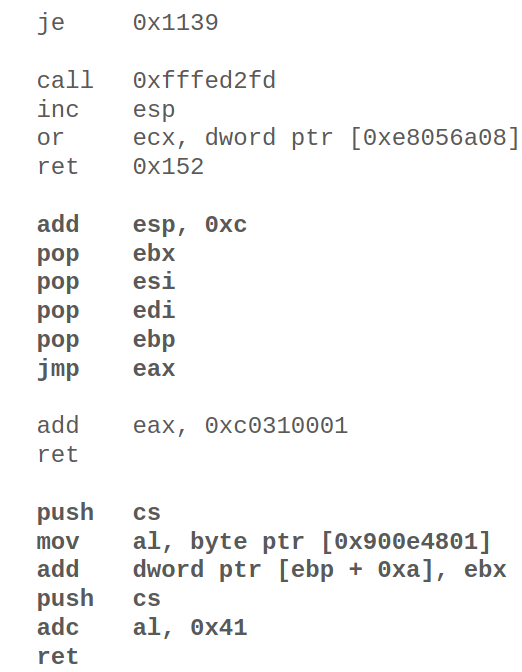}
         \caption{Selected Sample From Target Domain (vsftpd)}
         \label{fig:h2_showcase_2}
     \end{subfigure}

    \caption{Selected Samples for H2. Similar Two Samples.}
    \label{fig:h2_showcase}
\end{figure}

To verify \textbf{H2}, we want to find two similar instruction sequences, one from each domain, and see if their intermediate outputs are similar as well. The intuition is that since the model will remember the useful features, it can extract similar features from two similar instruction sequences from different domains. Figure \ref{fig:h2_showcase} shows two similar data samples (i.e. instruction sequences) from the two domains, respectively where the similar gadgets are in bold. We first show the similarity between the two gadget-chain snippets using semantic explanation. As shown in the Figure \ref{fig:h2_showcase}, both gadget-chain snippets are trying to first manipulate the stack for the next gadget (Write-What-Where gadgets), and then manipulate the \texttt{eax} register for system calls (Init system call gadgets). However, since it is from different programs, we can see that the actual instructions are different, but some common gadgets can be found.

Then, we use a quantified distance measure to show the similarity. First, the baseline distance is calculated by averaging the combination pairwise euclidean distance between source and target domain intermediate output from our trained transfer learning model. Note that different from what has been done in H1, this time all the intermediate outputs are from our trained transfer learning model. The baseline distance is 0.0141, and the distance between the intermediate output of the two code snippets in Figure \ref{fig:h2_showcase} is 0.0054. The distances show that the two code snippets selected have similar intermediate output. Also, if observed carefully, the gadgets from the proftpd seems a super set of the gadgets in vsftpd in this case, which corresponds to the observation stated in Section \ref{sec:improvement}: proftpd contains most of high quality features for ROP detection for other programs.

\section{Limitation and Conclusion}

Before the conclusion, we identify several limitations of our approach. First, although very few, minority class data samples are still needed for validation purposes. This could make our approach impractical if the minority class samples are completely unavailable or extremely rare. The assumption of our approach is that it is very difficult, but not impossible to generate benign samples in DeepReturn. Second, our approach requires high-quality source domain data. During the experiments, we observe that the quality of the source domain data can affect the performance substantially. Third, as illustrated in Section \ref{sec:improvement}, the selection of a source domain program is important to achieve good results. However, currently we do not have a method to determine what programs are good to serve as a source domain program.

In conclusion, this paper presents a transfer learning method to mitigate the imbalanced data issue in cybersecurity, using Return-Oriented Programming (ROP) payload detection as a case study. We propose a new domain adaptation based method to train a cyber-attack detection model using extremely imbalanced dataset; discuss the performance trade-offs of the proposed approach; and discuss the insights about how domain adaptation helps to achieve better results. Both strength and the limitation of our approach are discussed, and the FPR vs. DR trade-off is being identified.

\newpage
\bibliographystyle{unsrtnat}
\bibliography{references}  






\end{document}